\documentclass[iop]{emulateapj}
\usepackage{apjfonts}

\usepackage{epsfig}

\newcommand{\chandra}{\emph{Chandra}}
\newcommand{\xmm}{\emph{XMM-Newton}}
\newcommand{\xmms}{\emph{XMM}}

\newcommand{\s}{\ensuremath{\mbox{~s}}}
\newcommand{\ps}{\ensuremath{\s^{-1}}}

\newcommand{\km}{\ensuremath{\mbox{~km}}}
\newcommand{\Mpc}{\ensuremath{\mbox{~Mpc}}}
\newcommand{\pMpc}{\ensuremath{\Mpc^{-1}}}
\newcommand{\kmpspMpc}{\ensuremath{\km \ps \pMpc\,}}
\newcommand{\erg}{\ensuremath{\mbox{~erg}}}
\newcommand{\ergps}{\ensuremath{\erg \ps}}

\newcommand{\Ho}{\ensuremath{H_\mathrm{0}}}

\newcommand{\W}{\ensuremath{\mbox{~W}}}
\newcommand{\Hz}{\ensuremath{\mbox{~Hz}}}
\newcommand{\pHz}{\ensuremath{\Hz^{-1}}}
\newcommand{\WpHz}{\ensuremath{\W \pHz}}
\newcommand{\Sr}{\ensuremath{\mbox{~Sr}}}
\newcommand{\pSr}{\ensuremath{\Sr^{-1}}}
\newcommand{\WpHzpSr}{\ensuremath{\W \pHz \pSr}}

\begin{document}

\title{Heating the hot atmospheres of galaxy groups and clusters with cavities: the relationship between jet power and low-frequency radio emission}
\author{E. O'Sullivan\altaffilmark{1,2}, S. Giacintucci\altaffilmark{2,3,4},
  L.~P. David\altaffilmark{2}, M. Gitti\altaffilmark{2,5},
  J. M. Vrtilek\altaffilmark{2}, S. Raychaudhury\altaffilmark{1} and T.~J. Ponman\altaffilmark{1}}
\altaffiltext{1}{School of Physics and Astronomy, University of Birmingham, Birmingham, B15 2TT, UK, email: \textit{ejos@star.sr.bham.ac.uk}} 
\altaffiltext{2}{Harvard-Smithsonian Center for Astrophysics, 60 Garden
  Street, Cambridge, MA 02138, USA}
\altaffiltext{3}{INAF - Instituto di Radioastronomia, via Gobetti 101,
  I-40129, Bologna, Italy}
\altaffiltext{4}{Department of Astronomy, University of Maryland, College
  Park, MD 20742-2421, USA}
\altaffiltext{5}{Osservatorio Astronomico di Bologna - INAF, via Ranzani 1, I-40127, Bologna, Italy}

\shorttitle{Heating galaxy groups with cavities}
\shortauthors{O'Sullivan et al}

\begin{abstract}
  We present scaling relations between jet power and radio power measured
  using the Giant Metrewave Radio Telescope (GMRT), \chandra\ and \xmm, for
  a sample of 9 galaxy groups combined with the B\^{i}rzan et al.  sample
  of clusters. Cavity power is used as a proxy for mechanical jet power.
  Radio power is measured at 235~MHz and 1.4~GHz, and the integrated
  10~MHz-10~GHz radio luminosity is estimated from the GMRT 610-235~MHz
  spectral index. The use of consistently analysed, high resolution
  low-frequency radio data from a single observatory makes the radio powers
  for the groups more reliable than those used by previous studies, and the
  combined sample covers 6-7 decades in radio power and 5 decades in cavity
  power.  We find a relation of the form P$_{\rm jet}\propto\ $L$_{\rm
    radio}^{\sim0.7}$ for integrated radio luminosity, with a total scatter
  of $\sigma_{\rm Lrad}$=0.63 and an intrinsic scatter of $\sigma_{i,\rm
    Lrad}$=0.59. A similar relation is found for 235~MHz power, but a
  slightly flatter relation with greater scatter is found for 1.4~GHz
  power, suggesting that low--frequency or broad band radio measurements
  are superior jet power indicators. We find our low--frequency relations
  to be in good agreement with previous observational results. Comparison
  with jet models shows reasonable agreement, which may be improved if
  radio sources have a significant low--energy electron population. We
  consider possible factors which could bias our results or render them
  more uncertain, and find that correcting for such factors in those groups
  we are able to study in detail leads to a flattening of the P$_{\rm
    jet}$:L$_{\rm radio}$ relation.
\end{abstract}


\section{Introduction}
X--ray observations of clusters and groups over the past decade have
provided strong evidence that, despite central cooling times significantly
shorter than the Hubble time \citep[e.g.,][]{Sandersonetal06}, relatively
little gas actually cools below $\sim$0.5~keV
\citep{Petersonetal03,Kaastraetal04}. It is now widely accepted that
active galactic nuclei (AGN) in the central dominant ellipticals of these
systems can reheat the gas through a variety of mechanisms \citep[][and
references therein]{PetersonFabian06,McNamaraNulsen07}.

The most commonly observed evidence of interaction between AGN and the
intra-cluster or intra-group medium (IGM) is the presence of cavities,
formed when AGN jets inflate radio lobes, and identified from the resulting
X--ray surface brightness decrement. Cavities provide a relatively simple
method for estimating the power output of the jets, since the mechanical
energy required to expel the IGM can be estimated from the cavity volume
and surrounding pressure, and the timescale over which the cavity has
formed can be estimated from dynamical arguments. Cavities are expected to
heat the surrounding gas via the turbulent wake produced as they rise
buoyantly through the IGM \citep{Churazovetal01}. Studies of cavities have
shown that the energies required to create them are sufficient to suppress
cooling in systems across a wide range of mass scales
\citep{Birzanetal04,Dunnetal05,Raffertyetal06,Dunnetal10}.

The relationship between the mechanical power and radio emission of AGN
jets and lobes is of interest for two main reasons: Firstly because it
provides insight into the physical nature of the jet
\citep[e.g.,][]{Willottetal99}; Secondly because it allows estimation of
the energy available from AGN based on more easily acquired radio data
\citep[e.g.,][]{Bestetal07}.  \citet{Birzanetal04} determined the relation
between cavity power and 1.4~GHz radio power, using a sample dominated by
galaxy clusters. However, many cavities are undetected at 1.4~GHz since
radiative aging will cause higher frequency emission to fade fastest once
the jets cease to inject new plasma into the lobes.  \citet[][hereafter
B08]{Birzanetal08special} addressed the problem of these ghost cavities by
measuring the relation at 327~MHz, and using an estimate of the
integrated 10~MHz--10~GHz radio luminosity, both of which should be more
reliable, and both of which produced steeper relations.  \citet[][hereafter
C10]{Cavagnoloetal10special} extended the relation to lower jet and radio
powers by adding 21 giant ellipticals to the B08 sample, again finding a
steeper slope, but were hampered by the poor quality of available archival
low-frequency radio measurements.

We have compiled a sample of 18 galaxy groups, chosen to show signs of
AGN/IGM interactions, and observed both by the Giant Metrewave Radio
Telescope (GMRT) and by \chandra\ and/or \xmm\ \citep[][hereafter
G11]{Giacintuccietal11_special}. Of these groups, nine have cavities
(identified as decrements in surface brightness and in some cases as
temperature or abundance features) which are clearly correlated with radio
structures, and in this paper we add these to the B08 sample to examine the
relations between jet mechanical power and radio power. Our sample has
several advantages over previous studies: 1) Our low-frequency data were
acquired from a single observatory and are analysed uniformly, making both
flux density and spectral index measurements more reliable than is possible
for data collected from mixed archival sources; 2) We are able, for the
first time, to measure the integrated radio luminosity for a significant
number of low--radio--luminosity systems, as well as single-frequency
powers; 3) We have a closer correlation between radio and X--ray
morphologies in most cases, than is possible either at 1.4~GHz (where many
cavities are undetected) or with low-resolution low-frequency observations
(where unrelated sources may be difficult to remove); 4) Our groups have
low radio and cavity powers (typically P$_{\rm cav}\la 10^{44}$\ergps and
P$_{\rm 1400}\la 10^{24}$\WpHz), filling in a region of parameter space
sparsely populated in the B08 sample.

We describe our sample and analysis techniques in \S\ref{sec:obs}. Our
results and their relation to previous work are discussed in
\S\ref{sec:res}, and we present our conclusions in \S\ref{sec:conc}. A
$\Lambda$CDM cosmology with \Ho=70\kmpspMpc\ and
$\Omega_{M}$=1-$\Omega_{\Lambda}$=0.3 is adopted throughout the paper. The
radio spectral index $\alpha$ is defined as $S_\nu\propto\nu^{-\alpha}$,
where $S_\nu$ is the flux density at frequency $\nu$.

\section{Observations and Data Analysis}
\label{sec:obs}
The nine groups in our study were selected to have cavities and X--ray data
of sufficient quality to allow a reliable determination of the properties
of the IGM. Table~\ref{tab} lists the properties of the sample.  Images and
an in-depth description of the radio properties of the groups are presented
in G11. We do not include the giant sources NGC~315, NGC~383 and NGC~7626,
since the available data do not allow clear identification of cavities.
NGC~315 and NGC~383 extend outside the \xmms\ field of view, and NGC~7626
is in a merging group \citep{Randalletal09} where determination of IGM
properties and gravitating mass is unreliable.  NGC~1407 is not included
since the cavity tentatively identified by \citet{Dongetal10} is on a much
smaller scale than the radio emission, and its identification and size vary
with image processing.  NGC~741, where a ghost cavity has been previously
identified \citep{Jethaetal08} is excluded since the visible radio emission
appears to be unrelated to the cavity, and is contaminated by a second AGN
within the group. UGC~408 was not listed as a cavity system in G11, as no
detailed examination of its structure has yet been published.  However, it
is included in this sample, since its radio source appears to have cleared
an elliptical cavity in the group centre; it was also included in the C10
sample. A more detailed description of IGM structures in our full sample of
18 groups will be presented in a later paper.

In three cases we had already published individual studies of the cavities;
AWM~4 \citep{OSullivanetal10a}, HCG~62 \citep{Gittietal10} and NGC~5044
\citep{Davidetal09}. For these groups, we used the gas properties
determined from these prior analyses. We note that for HCG~62 the gas
temperature and density profiles were extracted from an \xmm\ observation;
in all other cases \chandra\ data were used.

For the remaining systems, we determined the gas properties from the
longest available \chandra\ observations. Table~\ref{tab} provides basic
information on each dataset. Observations were downloaded from the
\chandra\ archive and reprocessed using \textsc{ciao} 4.2 and
\textsc{caldb} 4.3.0 following the standard techniques described in the
\chandra\ analysis
threads\footnote{http://asc.harvard.edu/ciao/threads/index.html} and
\citet{OSullivanetal10a}. Point sources were identified using the
\textsc{wavdetect} task and, with the exception of sources coincident with
the group-central AGN, removed. As the groups typically fill the field of
view, background spectra were drawn from the standard set of ACIS blank sky
background events files, scaled to produce the same 9.5-12.0 keV count rate
as that in the target observation. Very faint mode screening was applied to
target and background observations where appropriate.

\begin{deluxetable*}{lccccccccc}
\label{tab}
\tablewidth{0pt}
\tablecaption{\label{tab} Summary of observations and data for each group}
\tablehead{
\colhead{Group} & \colhead{Instrument} & \colhead{ObsID} &
\colhead{t$_{exp}$} & \colhead{D} & \colhead{$\alpha^{610}_{235}$} & 
\colhead{P$_{1400}$} & \colhead{P$_{235}$} & \colhead{L$_{\rm radio}$} &
P$_{\rm cav}$$^d$ \\  
\colhead{} & \colhead{} & \colhead{} & \colhead{(ks)} & \colhead{(Mpc)} &
\colhead{} & \colhead{(10$^{24}$\WpHz)} & \colhead{10$^{24}$\WpHz)} &
  \colhead{(10$^{42}$\ergps)} & \colhead{(10$^{42}$\ergps)} 
}
\startdata
AWM~4    & ACIS-S & 9423  & 74.5 & 128.7 & 0.67 & 1.41$\pm$0.07     & 6.31$\pm$0.50     & 0.14$\pm$0.02        & 53.62$^{+29.35}_{-9.05}$\\[+1mm]
HCG~62   & ACIS-S & 10462 & 67.1 & 57.8  & 1.15 & 0.0021$\pm$0.0001 & 0.018$\pm$0.001   & 0.00029$\pm$0.00003 & 3.49$^{+0.91}_{-1.66}$\\
         & XMM    & 050478501 & 122.5 & - & - & - & - & - & - \\
UGC~408  & ACIS-S & 11389 & 93.9 & 61.9  & 0.53 & 0.83$\pm$0.04     & 2.57$\pm$0.21     & 0.071$\pm$0.008     & 3.53$^{+2.69}_{-2.75}$\\[+1mm]
NGC~507  & ACIS-I & 2882  & 43.5 & 69.3  & 1.14 & 0.060$\pm$0.003   & 0.68$\pm$0.05     & 0.011$\pm$0.001     & 20.68$^{+28.72}_{-5.58}$\\[+1mm]
NGC~4636 & ACIS-I & 3926  & 74.6 & 13.2  & 0.61 & 0.0017$\pm$0.0001 & 0.0053$\pm$0.0004 & 0.00013$\pm$0.00002 & 0.28$^{+0.11}_{-0.04}$\\[+1mm]
NGC~5044 & ACIS-S & 9399  & 82.7 & 38.2  & $>$1.90$^a$ & 0.0065$\pm$0.0003 & 0.042$\pm$0.003   & 0.0019$\pm$0.0002   & 2.88$^{+2.55}_{-0.45}$\\[+1mm]
NGC~5813 & ACIS-S & 9517  & 98.8 & 26.8  & 0.94$^b$ & 0.0014$\pm$0.0001 & 0.0079$\pm$0.0006 & 0.00013$\pm$0.00002 & 1.64$^{+0.29}_{-0.20}$\\[+1mm]
NGC~5846 & ACIS-I & 7923  & 90.0 & 24.3  & 0.65$^c$ & 0.0015$\pm$0.0001 & -                 & 0.00011$\pm$0.00001 & 1.71$^{+0.50}_{-0.44}$\\[+1mm]
NGC~6269 & ACIS-I & 4972  & 39.6 & 142.9 & 0.73 & 0.141$\pm$0.007   & 0.65$\pm$0.05     & 0.013$\pm$0.002     & 10.02$^{+2.20}_{-2.80}$
\enddata
\tablecomments{ $^a$ Only a lower limit on the 235-610~MHz spectral index,
  $\alpha^{610}_{235}$, could be estimated for NGC~5044, as extended
  structures at 235~MHz are undetected at higher frequencies (see G11).  $^b$
  NGC~5813 spectral index measured between 235~MHz and 1.4~GHz.  $^c$
  NGC~5846 spectral index measured between 610~MHz and 1.4~GHz. $^d$ Cavity power estimated based on the buoyancy timescale and including uncertainties on
volume, pressure, and timescale.}
\end{deluxetable*}

Spectra were extracted from circular annuli and a deprojected absorbed
APEC model \citep{Smithetal01} was fitted to each set of spectra using
\textsc{xspec} v12.6.0k. The absorbing column was fixed at the galactic
value determined from the survey of \citet{Kalberlaetal05}. Redshifts and
adopted distances for each group are given in G11.
Energies below 0.5~keV and above 7~keV were ignored during fitting, so as
to minimise calibration and background uncertainties.

Jet powers were assessed using the standard approach of assuming that the
mechanical power of the jet can be approximated as the energy of the
detected cavities averaged over some timescale
\citep[e.g.,][B08]{Birzanetal04}. We define the energy of each cavity to be
4$pV$, the enthalpy of a cavity filled with a relativistic plasma. Cavity
sizes are defined by matching an ellipse to their apparent shape, and
assuming a line-of-sight depth equal to the minor axis. Uncertainties on
the volume are estimated by assuming a minimum depth of half this value,
and a maximum equal to the major axis. The uncertainty on the volume is, in
most cases, the largest contributor to the uncertainty on the energy of
each cavity and the total uncertainty on the cavity power.  Outburst
timescales are typically estimated from the buoyant rise time of the
cavities \citep[$t_{buoy}$,][]{Churazovetal01}, but we also estimate the
sonic timescale (i.e., the travel time between the AGN and farthest point
of the cavity at the sound speed) and refill timescale of the cavities.
Our cavity power estimates assume the buoyancy timescale, to ensure that
they are directly comparable with previous studies. The uncertainty on this
estimate includes the uncertainties on volume, pressure, and t$_{buoy}$.
These values are used when fitting the relations between cavity power and
radio power. The additional uncertainty associated with the other timescale
estimates is indicated when we plot these relations, but is not included in
the fitting process.

The analysis of our GMRT radio data is described in G11. Most of the groups
in the sample were observed at both 235~MHz and 610~MHz, but of the cavity
systems included in this study, NGC~5813 was only observed at 235~MHz and
NGC~5846 only at 610~MHz.  Radio powers at each frequency, P$_\nu$, are
estimated as P$_\nu$=4$\pi$D$_L^2$(1+z)$^{\alpha-1}$S$_\nu$, where
S$_\nu$ is the flux density at frequency $\nu$, $\alpha$ the spectral
index, z is the redshift and D$_L$ is the luminosity distance to the
source. Radio powers at 1.4~GHz were also estimated from flux densities
mainly derived from the NRAO VLA Sky Survey \citep[NVSS,][]{Condonetal98}.

Since it is desirable to compare our 235~MHz GMRT measurements with the
327~MHz VLA radio powers used by B08, we have used the spectral indices
given in B08 and G11 to correct each set of flux densities to the other
band, and determined the best-fitting relationship between cavity power and
radio power for both frequencies. It should be noted that while G11
determined 235-610~MHz spectral indices for the majority of their sample,
only a lower limit on the spectral index could be determined for NGC~5044
($\alpha>1.9$), and no index could be determined for NGC~5813 or NGC~5846,
since they were only observed at one frequency. When correcting those data
to 327~MHz, we adopt a spectral index of $\alpha$=1.9 for NGC~5044, the
235~MHz-1.4~GHz spectral index, $\alpha$=0.94, for NGC~5813, and the
610~MHz-1.4~GHz spectral index, $\alpha$=0.65, for NGC~5846.

We also estimate the integrated radio luminosity of our groups in the
10~MHz-10~GHz band, assuming a powerlaw spectrum and using the same
spectral indices. These are not perfectly comparable with the integrated
radio power estimates of B\^{i}rzan et al., which are based on model fits
to flux densities measured at 3-8 frequencies for each system. However, the
inaccuracy introduced by our use of a simple spectral index is unlikely to
be significant compared to the large uncertainties in cavity power.
Low-frequency spectral index measurements such as ours should not be
strongly affected by radiative losses unless the break frequency is below
our lowest measurement frequency (235~MHz). It seems unlikely that any of
our systems are so old. Two groups in the sample, NGC~507 and AWM~4, have
unusually high radiative age estimates
\citep{Murgiaetal11,Giacintuccietal08} but break frequencies of
300-450~MHz, above our lower frequency bound. If the spectral indices are
accurate, then we are unlikely to over-estimate the steepness of the
spectrum and the power at low frequencies, as would probably be the case if
high-frequency spectral indices were used.  Detailed modelling of the radio
spectra for our group sample and revised estimates of integrated radio
power will be presented in a forthcoming paper.

\section{Results and Discussion}
\label{sec:res}
\begin{figure*}[t]
\includegraphics[width=\columnwidth,bb=20 200 570 750]{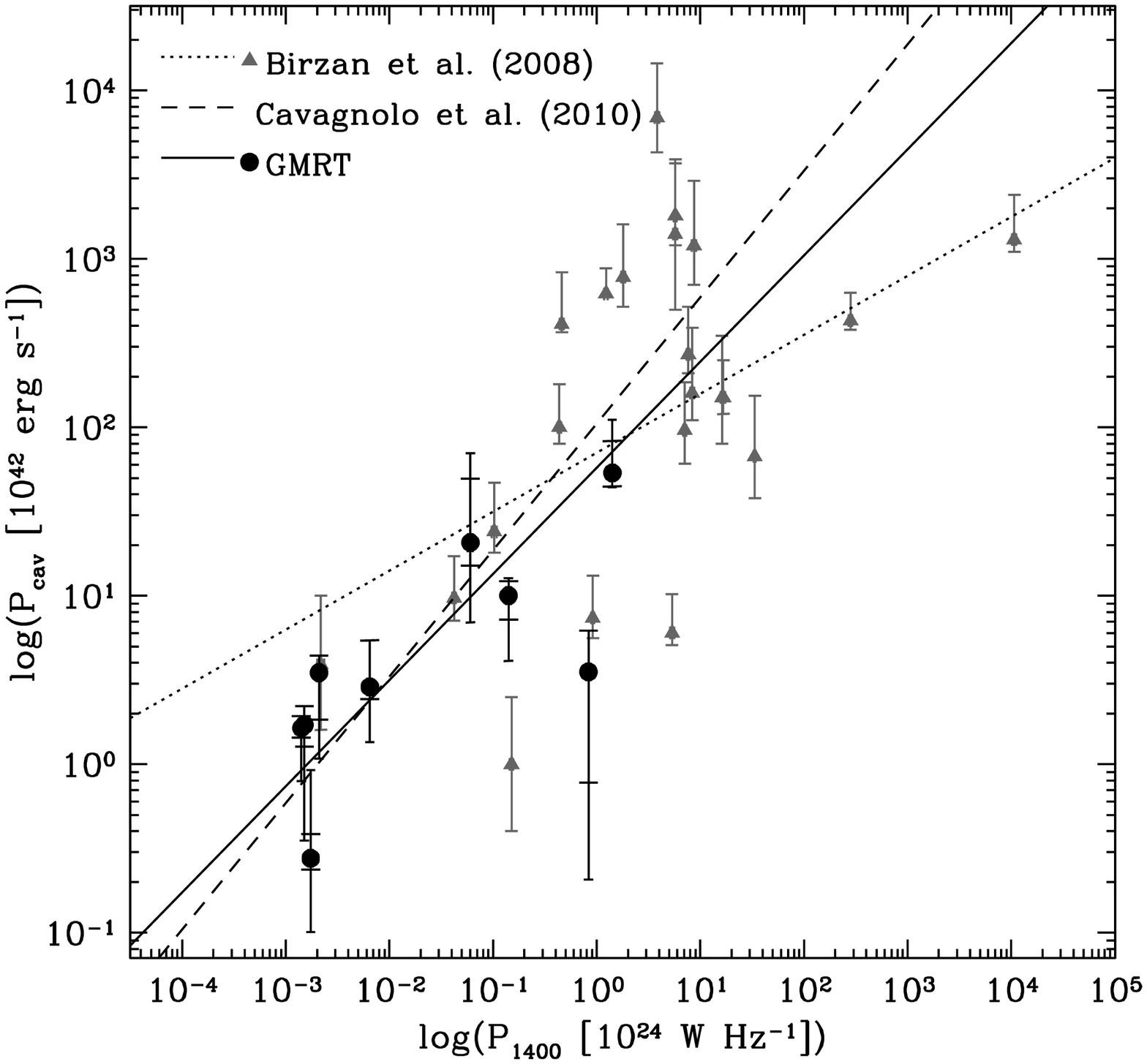}
\includegraphics[width=\columnwidth,bb=20 200 570 750]{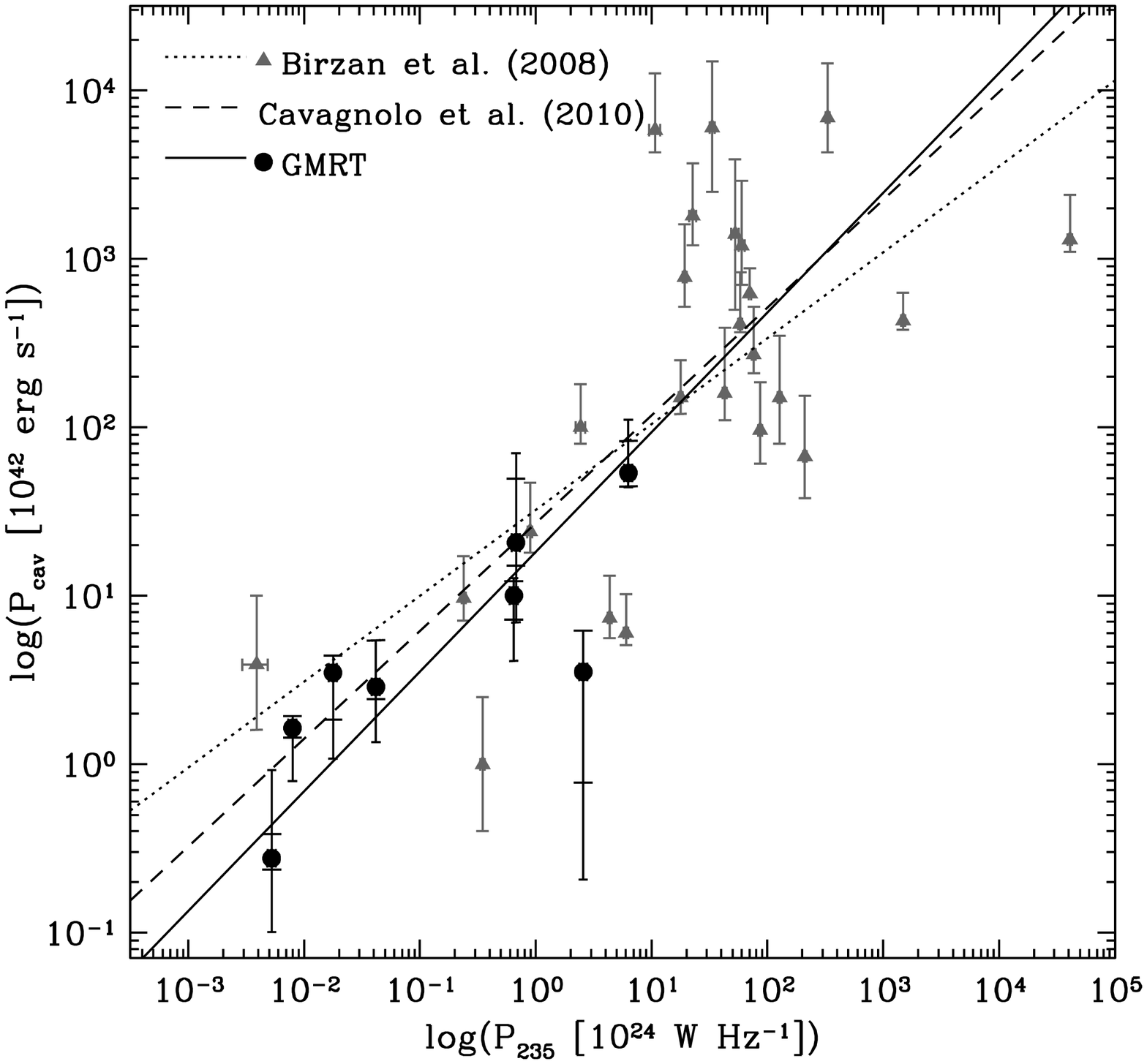}
\caption{\label{fig:1400} Cavity power vs. radio power at 1.4~GHz (left)
  and 235~MHz (right). Systems in the sample of B08 are marked by grey
  triangles, members of our groups sample by black circles.  1$\sigma$
  uncertainties on radio power and cavity power (calculated using the
  buoyant rise time of the cavities) are indicated by error bars. In many
  cases the uncertainty on the radio luminosity is smaller than the data
  point. For our groups, additional narrow-width error bars indicate the
  1$\sigma$ uncertainty range allowing for alternate measures of cavity age
  (the sonic and refill timescales).  The B\^{i}rzan et al. point with the
  lowest radio power is HCG~62, which is also included in our GMRT groups
  sample. This point is included for comparison, but was excluded from our
  analysis.  The solid fit line indicates our BCES regression fit to the
  data points. The dotted line indicates the relation found by B08, the
  dashed line the relation found by C10. For the 235~MHz relation, we have
  used the spectral indices given by B08 to correct their 327~MHz data to
  235~MHz, so as to allow a direct comparison with our data.  The
  normalizations of the B\^{i}rzan or Cavagnolo fit lines are not corrected
  for frequency differences, but the lines are indicative of the relative
  gradient of the different fits.  }
\end{figure*}

\begin{figure}
\includegraphics[width=\columnwidth,bb=20 200 570 750]{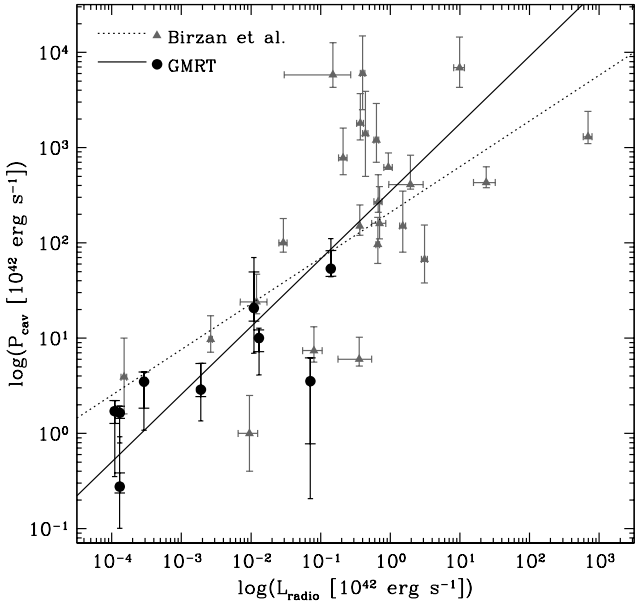}
\caption{\label{fig:Lrad} Cavity power vs. integrated 10~MHz-10~GHz radio power. Symbols are the same as in Fig.\ref{fig:1400}.}
\end{figure}

Figure~\ref{fig:1400} shows cavity power vs. radio power at 1.4~GHz and
235~MHz for our nine groups and the 24 systems described by B08. At both
frequencies, the scatter among the points is considerable. The effect of
radiative aging on the radio sources causes preferential fading of emission
at higher frequencies. We therefore expect the low-frequency measurements
to provide a more accurate estimate of the true radio power in older
systems. Figure~\ref{fig:Lrad} shows the relationship between cavity power
and the integrated radio luminosity in the 10~MHz-10~GHz band. The
integrated luminosity should be a superior measure of the radiative power
compared to estimates at a single frequency, since it accounts for
variations in spectral index between sources. However, we still see a large
degree of scatter amongst the data points.

We used the bivariate correlated error and intrinsic scatter (BCES)
algorithm \citep{AkritasBershady96} to perform linear regression fits to
the data, determining the best fitting power-law relationship between
cavity power, P$_{\rm cav}$, and the radio powers at each frequency
(P$_{1400}$, P$_{235}$, P$_{327}$) or the integrated radio
luminosity L$_{\rm radio}$. Using the orthogonal BCES regression to the
parameters in log space, the best-fit relations are:

\begin{equation}
{\rm log}\; {\rm P}_{\rm cav} = 0.71\; (\pm0.11)\; {\rm log}\; {\rm P}_{235} + 1.26\; (\pm0.12)
\end{equation}
\begin{equation}
{\rm log}\; {\rm P}_{\rm cav} = 0.72\; (\pm0.11)\; {\rm log}\; {\rm P}_{327} + 1.38\; (\pm0.12)
\end{equation}
\begin{equation}
{\rm log}\; {\rm P}_{\rm cav} = 0.63\; (\pm0.10)\; {\rm log}\; {\rm P}_{1400} + 1.76\; (\pm0.15)
\end{equation}
\begin{equation}
{\rm log}\; {\rm P}_{\rm cav} = 0.71\; (\pm0.11)\; {\rm log}\; {\rm L}_{\rm radio} + 2.54\; (\pm0.21)
\end{equation}

where P$_{\rm cav}$ and L$_{\rm radio}$ are in units of 10$^{42}$\ergps,
and P$_{235}$, P$_{327}$ and P$_{1400}$ in units of
10$^{24}$\WpHz.

We estimated the total scatter in the data using the technique described in
\citet{Prattetal09}, which is based on weighted estimates of the orthogonal
distance between the data points and best fit lines. We find the total
scatter for the radio power relations to be $\sigma_{1400}$=0.68 dex and
$\sigma_{235}$=0.62 dex, while the integrated radio luminosity relation has
a scatter of $\sigma_{\rm Lrad}$=0.63. The intrinsic scatter in the data
can be estimated by subtracting the contribution expected from the
statistical errors. We found that the scatter about the radio power
relations was $\sigma_{i,1400}$=0.65 dex and $\sigma_{i,235}$=0.58 dex, and
the scatter about the integrated radio luminosity relation was
$\sigma_{i,\rm Lrad}$=0.59 dex.

These relations can be compared with those derived by B08,
including the correction to the 327~MHz relation given by
\citet{Birzanetal10} and using the integrated radio luminosity from their
sources as a whole rather than from the lobes alone:
\begin{equation}
{\rm log}\; {\rm P}_{\rm cav} = 0.51\; (\pm0.07)\; {\rm log}\; {\rm P}_{327} + 1.51\; (\pm0.12)
\end{equation}
\begin{equation}
{\rm log}\; {\rm P}_{\rm cav} = 0.35\; (\pm0.07)\; {\rm log}\; {\rm P}_{1400} + 1.85\; (\pm0.10)
\end{equation}
\begin{equation}
{\rm log}\; {\rm P}_{\rm cav} = 0.49\; (\pm0.07)\; {\rm log}\; {\rm L}_{\rm radio} + 2.32\; (\pm0.09)
\end{equation}

However, these relations were estimated using an ordinary least-squares
regression, minimising the distance between fit line and data points in the
P$_{\rm cav}$ axis only. Refitting these relations using the orthogonal
BCES regression steepens them and increases the uncertainties, giving
slopes of 0.67$\pm$0.19 at 327~MHz, 0.57$\pm$0.17 at 1.4~GHz and
0.68$\pm$0.19 for the integrated radio power. HCG~62, the only group in the
B08 sample, has a strong influence. Excluding this system, fits to the B08
dataset have gradients of 0.81$\pm$0.30 at 327~MHz, 0.62$\pm$0.28 at
1.4~GHz, and 0.78$\pm$0.30 for the integrated radio power. This steepening,
and the large uncertainties on the gradients, emphasize the need for a
large dynamic range if gradients are to be measured accurately from data
with such a large degree of scatter.

B08 also found steeper relations when fitting only those systems where the
radio emission fills the cavities seen in the X-ray, excluding systems with
ghost cavities. A comparison of radio and cavity power on this basis is
likely to be more reliable, since the radio emission is clearly associated
with the lobes which inflated the cavity. Our relations have gradients
which are steeper than the published B08 relations (though consistent with
them at the 1.6-2.3$\sigma$ significance level), and in agreement with
those derived from the B\^{i}rzan et al.  sample using orthogonal
regression.  The total scatter about the B08 radio luminosity
relation is $\simeq$0.83 dex, suggesting that our groups are more closely
clustered about the relation.

\citet{Cavagnoloetal10special}, using orthogonal BCES regression, found the
following relations between radio power and cavity power:

\begin{equation}
{\rm log}\; {\rm P}_{\rm cav} = 0.64\; (\pm0.09)\; {\rm log}\; {\rm P}_{200-400} + 1.54\; (\pm0.12)
\end{equation}
\begin{equation}
{\rm log}\; {\rm P}_{\rm cav} = 0.75\; (\pm0.14)\; {\rm log}\; {\rm P}_{1400} + 1.91\; (\pm0.18)
\end{equation}

Unsurprisingly, given the large overlap between our sample and that of C10
(The C10 sample of 21 ellipticals includes 8 of our 9 groups, and both sets
of fits have the B08 sample in common) their best-fitting relations are
very similar to those we derive, with gradients in agreement to within the
1$\sigma$ uncertainties at both 1.4~GHz and at lower frequencies. C10 drew
their low frequency radio measurements from the CATS database
\citep{Verkhodanovetal97}, a compilation of data from numerous radio
catalogues. We expect there to be significant variations in sensitivity and
resolution among measurements from different observatories and surveys. Our
GMRT 235~MHz were observed and processed uniformly, and we therefore expect
our data points to be more reliable. Comparison of our 235~MHz radio powers
with the 200-400~MHz powers of C10 suggests that agreement is fairly good
for the most radio luminous systems (P$_{235}\ga 10^{23}$\WpHz). However,
for the fainter sources, disagreements of a factor of $\sim$10 are typical.
This may be explained by issues such as the inclusion of unresolved
background sources and the non-detection of faint structures. The total
scatter about the C10 relations is comparable with ours at low frequencies
($\sigma_{200-400}$=0.61) and slightly larger at 1.4~GHz
($\sigma_{1400}$=0.78).

\subsection{Comparison with theoretical models}
\citet{Willottetal99} determined a theoretical relation between radio
luminosity and jet power which is now widely used to estimate the kinetic
energy output of AGN \citep[e.g.,][]{Hardcastleetal07,CattaneoBest09}. 
This model takes the form

\begin{equation}
{\rm P}_{\rm jet} = 3\times10^{38}f^{3/2}{\rm P}^{6/7}_{151} \,\,\,\,\,\,\,  {\rm W},
\end{equation}

where P$_{\rm jet}$ is the total jet power, P$_{151}$ is its observed
151~MHz radio power in units of 10$^{28}$\WpHzpSr, and the factor $f$
includes a variety of unknown factors affecting the normalisation of the
relation. Under the assumption that the spectra of radio sources are simple
powerlaws, the normalisation of the model is dependent on frequency, but
its gradient is not. Since P$_{\rm cav}$ is a proxy for the total jet
power, it is straightforward to compare the model gradient with those of
measured P$_{\rm cav}$:P$_{\rm radio}$ relations. C10 compared their best
fitting P$_{\rm cav}$:P$_{1400}$ relation with the model, finding that its
gradient of $\sim$0.86 agreed with their measured gradient within errors.
At lower frequencies, which should be more reliable, the model agrees with
the C10 P$_{\rm cav}$:P$_{200-400}$ relation within $\sim$2.5$\sigma$
confidence bounds.  The normalisation of the model is dependent on a number
of factors, including $k$, the ratio of energy in non-radiating particles
to relativistic electrons. C10 found that $k$ values of tens to thousands
were required to bring the model normalisation into agreement with their
data. Ignoring their relative normalisations, the gradient of our 235~MHz
relation would agree with the model within 1.5$\sigma$, and our 1.4~GHz
relation within 2.5$\sigma$.

The \citet{Willottetal99} model is based on relatively straightforward
synchrotron physics, but includes a number of important assumptions.  The
jet power is proportional to the energy density of the relativistic plasma
assuming a minimum energy magnetic field, $u_{me}$, following the relation
P$_{jet} \propto u_{me}^{3/2} \propto B_{me}^3$, where $B_{me}$ is the
minimum energy magnetic field.  Willott et al. define $B_{me}$ in terms of
observable radio properties following the method given by \citet{Miley80},
which can be simplified to
$B_{me}\propto(\nu_2^{1/2-\alpha}-\nu_1^{1/2-\alpha})^{2/7}$, where $\nu_1$
and $\nu_2$ are frequencies defining the observable radio band
(10~MHz--100~GHz for Willott et al.) and $\alpha$ is the spectral index.
However, this assumes that the electron population in the radio-emitting
plasma only contains particles with energies such that they radiate in this
observable band, and Willott et al.  note that the jet power will depend
critically on the choice of low frequency cutoff.

Electrons with energies too low to produce observable emission may
contribute a significant fraction of the energy in radio jets and lobes,
since if they are present they are likely to make up the majority of the
electron population.  Many observational studies define the range of
electron energies in terms of a range of Lorentz factors, typically with a
low energy cutoff of $\gamma_{\rm min}$=10 or 100. In a magnetic field of
$\sim$1$\mu$G, such electrons will produce synchrotron radiation at
$\sim$0.5-50~kHz, well below the observable band.
\citet{WorrallBirkinshaw06} provide a relation between $B_{me}$ and
observable radio parameters using Lorentz factors rather than frequencies,
which takes the form $B_{me}\propto(\gamma_{\rm
  max}^{1-2\alpha}-\gamma_{\rm min}^{1-2\alpha})^{1/(\alpha+3)}$. If we use
this definition to determine $u_{me}$ and therefore P$_{\rm jet}$, the
dependence of the gradient of the Willott et al. model on the spectral
index becomes clear, giving a relation of the form:

\begin{equation}
{\rm P}_{\rm jet} \propto {\rm P}^{3/(\alpha+3)}_{\rm radio}.
\end{equation}

This reduces to a gradient of 6/7 for $\alpha=0.5$. However, the measured
610-235~MHz spectral indices of the G11 sample cover
the range 0.53-1.44. Excluding systems for which only limits are available,
the mean spectral index is $\alpha^{610}_{235}=0.95$, implying a
P$_{\rm cav}$:P$_{\rm radio}$ gradient of 0.76. This is
considerably closer to our best measured gradients of $\sim$0.7 for 235~MHz
power and integrated radio luminosity. A spectral index of 0.8, often used
as a typical value for extragalactic sources \citep{Condon92} would give a
model gradient of 0.79.  Our best fitting P$_{\rm cav}$:L$_{\rm radio}$
relation gradient of 0.71 would suggest $\alpha\sim1.2$.

Willott et al. developed their model to examine the relationship between
low-frequency radio power and narrow line optical luminosity in a sample of
7C and 3CRR radio sources. They found no correlation between 151~MHz
spectral index and the residual from their best fitting radio to optical
relation. Such a correlation would be expected if jet power varied strongly
with spectral index. They argue that this suggests that low-energy
electrons make only a minimal contribution to jet power. However, the
scatter in their dataset is large (as it is in ours) and their sample
contains a wide variety of source types, for which spectral index may be
dominated by emission from different physical regions. FR-II radio galaxies
are also common in their sample, while our groups host only moderately
powerful FR-I sources and a few FR-I/FR-II transition systems. It therefore
seems possible that the slope of the relation could be steeper.  Such a
change in gradient may have implications for models which have used the
Willott et al. relation to estimate the energy output of the population of
AGN from their radio luminosity function, with a shallower gradient
implying greater mechanical power available from lower luminosity jets.

\subsection{Uncertainties and potential biases}
Several factors could affect our estimates of the 4$pV$ cavity
power, or the accuracy of these estimates as a proxy for the mechanical
power of the jets. These include:

\begin{itemize}
\item Cavity volume. Identification and characterisation of the cavity is
  inherently subjective, and dependent on the quality of the X--ray data,
  the angular and physical sizes of the cavity, its position in the group
  and other factors. Using HCG~62, which has several \chandra\ and \xmm\
  datasets with a range of exposures available, we tested the effects of
  performing independent spectral deprojections and having different
  researchers estimate cavity size. We found differences between estimates
  of up to a factor of 2 ($\sim$0.3 dex) in total cavity power, despite the
  relatively simple morphology of the cavities.
\item Cavities at large radii are significantly more difficult to identify,
  owing to the decrease in X--ray intensity with radius. In both HCG~62 and
  NGC~5044, 235~MHz radio maps reveal lobes at large radii, beyond the
  cavities identified in the X--ray \citep{Davidetal09,Gittietal10}. In
  both cases there is some evidence of the presence of a cavity in the
  X--ray \citep[e.g., a low abundance region suggesting multi-phase gas
  coincident with the detached lobe in NGC~5044][]{Davidetal11}, and if
  such cavities exist and have sizes similar to the radio lobes, this would
  increase our estimate of the AGN power by a factor of $\sim$2-10.
  Multiple cavity pairs in individual groups also give some idea of the
  range of cavity powers. In both NGC~5044 and NGC~5813, cavity powers for
  individual cavities vary by factors of up to 10.
\item Very old or young cavities are unlikely to be detected.  Young, small
  cavities would have a minimal X--ray surface brightness decrement. Their
  apparent cavity powers would also be small, but since they would likely
  be highly over-pressured, cavity power would underestimate jet power.
  Very old cavities would be expected to have risen to very large radii or
  to have already broken up into smaller structures; in either case the
  surface brightness decrements would be small. Their cavity powers would
  be low, owing to their long timescales.  Non-detection of such cavities
  in groups could bias the P$_{\rm cav}$:P$_{\rm radio}$ relations to
  steeper gradients, but it is unclear how common such cavities are, or
  whether similar biases might also affect galaxy clusters.
\item Shocks driven by the AGN outburst may contain a large fraction of the
  energy released. Inclusion of the shocks detected in HCG~62 and NGC~5813
  \citep{Gittietal10,Randalletal11} would increase our estimate of the AGN
  power output by a factor $\sim$10. Given the difficulty of detecting
  shocks (deep \chandra\ observations are generally required) we cannot
  know whether we are missing a significant energy contribution in other
  groups.
\item Uncertainties in outburst timescale will also affect power estimates.
  We have used the buoyancy timescale and its uncertainties when fitting
  the P$_{\rm cav}$:P$_{\rm radio}$ relations, and included additional
  error bars to show the larger uncertainty range associated with different
  dynamical timescale estimates, but even these may not be accurate,
  particularly for systems which have expanded supersonically for a
  significant fraction of their lifespan (e.g., UGC~408). Very old sources
  may also provide poor estimates if the buoyant velocity overestimates
  their true rate of rise; radiative ages based on synchrotron losses can
  exceed the the dynamical timescales in such systems \citep[e.g., in AWM~4
  and NGC~507,][]{OSullivanetal10a,Murgiaetal11}.
\item Filling factors of less than unity for radio lobes could also render
  volume estimates for cavities inaccurate, affecting both the energy and
  dynamical timescales of the outburst. Our study of AWM~4 suggested that
  the lobes may have a filling factor as low as $\phi$=0.2
  \citep{OSullivanetal10a}.
\item Jet orientation could affect our estimate of the position of cavities
  in the IGM, with jets close to the line of sight producing cavities which
  appear to be at smaller radii than is the case. This will lead to
  underestimates of cavity enthalpy and outburst timescale. Simulations
  suggest that cavity powers will typically be within a factor of 3 of the
  true value \citep{Mendygraletal11}.
\item AGN--driven ``weather'', turbulent motions in the IGM induced by AGN
  jets, could affect the position of cavities. NGC~5044 may provide an
  example, with several small cavities found at similar radii in the group
  core \citep{Davidetal09}. Timescales for such cavities are probably
  underestimated, leading to overestimates of cavity power.
\end{itemize}

For our sample of groups, it appears that the effect of such biases may be to
steepen the relation between cavity power and radio power. Correcting the
cavity power estimates for HCG~62, NGC~5044, NGC~5813 and AWM~4 to include
contributions from the probable outer cavities and shocks, and the low
filling factor of the AWM~4 radio lobes, we find that the gradient of the
P$_{\rm cav}$:L$_{\rm radio}$ relation flattens to 0.62$\pm$0.12. This is
still consistent, within uncertainties, with our initial fit, but we note
that these are perhaps the four most carefully studied systems in our
sample, and deeper observations of other systems would likely lead to
similar corrections. For the systems we can study in detail, corrections
which reduce cavity power tend to be small, whereas the effect of including
shocks and possible additional cavities can increase the estimated AGN
power by much larger factors.
  
The impact of such biases on galaxy clusters is less clear. These are
generally more X--ray luminous systems with larger cores, hosting more
powerful radio sources, which might suggest that cavities will be more
easily detected. The intra-cluster medium is also likely to be able to
confine more powerful AGN outbursts, which in groups might simply tunnel
out to large radii where any cavities would be difficult to detect (e.g.,
the 'poorly--confined'' class of objects in the C10 sample). However, weak
shocks may be more easily detected in galaxy groups, since the emission
lines produced by gas at $\sim$1~keV significantly improve our ability to
accurately measure temperature. Resolution may also be a problem for more
distant clusters, producing a bias toward the detection of large cavities
associated with more powerful AGN outbursts. A statistical approach to the
effects of such biases, based on simulations, would perhaps provide insight
into this problem.

\section{Conclusions}
\label{sec:conc}
We have estimated the relations between mechanical jet power and radio
luminosity, adding nine groups selected from the sample of G11 and observed
with the GMRT and \chandra\ or \xmm\ to the B08 sample, which consists
primarily of galaxy clusters. We find P$_{\rm cav}$:P$_{\rm radio}$
relations with gradients of $\sim$0.7 for both the low-frequency radio
power (235 or 327~MHz) and the integrated radio luminosity, with total
scatters of $\sigma_{i,\rm Lrad}$=0.59 and $\sigma_{i,235}$=0.58 dex. The
1.4~GHz relation is somewhat flatter (gradient $\sim$0.6) and has a
slightly larger scatter, $\sigma_{i,1400}$=0.65 dex.  In agreement with
previous studies, this suggests that low-frequency and broad band radio
measurements are superior indicators of cavity power. The weaker
correlation between 1.4~GHz radio structure and the cavities identified
from the X--ray makes high frequency emission a poor choice for such
studies. Our fitted slopes are significantly steeper than those found by
B08, but this is unsurprising since the B08 relations were determined using
a different regression technique, which will tend to produce a shallower
gradient. Using the same BCES orthogonal regression used for the fits to
the combined dataset brings the B08 relations into agreement with our
results. Our P$_{\rm cav}$:P$_{235}$ relation has a similar gradient to the
P$_{\rm cav}$:P$_{200-400}$ relation of C10. This is expected, given the
overlap between samples, but direct comparison of the radio powers suggests
that our GMRT measurements are more reliable for low-power radio galaxies.

Our P$_{\rm cav}$:P$_{235}$ relation is somewhat flatter than the widely
used \citet{Willottetal99} model of jet mechanical and radio power, though
they agree within 1.5$\sigma$ uncertainties. We find that inclusion of
electrons with low Lorentz factors (which cannot be directly observed)
could change the gradient of the Willott model, making it dependent on the
radio spectral index. In this case, using the mean spectral index of the
G11 sample, the model would agree more closely with our observed relation.
A variety of factors could bias or increase the uncertainty of our
measurements, and we conclude that at least for galaxy groups these may
have a serious impact on our cavity power estimates. Correcting for these
factors, in those groups where the quality of radio and X--ray data is
sufficient to allow detailed study, produces a flatter P$_{\rm
  cav}$:P$_{\rm radio}$ relation. However, it is unclear whether this would
be the case in all groups, or in galaxy clusters, and simulations of AGN
feedback across a wide range of mass scales and outburst powers are
probably required to resolve this question.

\acknowledgments We thank the anonymous referee for their useful
suggestions. We thank the staff of the GMRT for their help during
observations. GMRT is run by the National Centre for Radio Astrophysics of
the Tata Institute of Fundamental Research. We also acknowledge support
from Chandra grant AR1-12014X. EJOS thanks A.~J.~R. Sanderson for useful
discussions and acknowledges the support of the European Community under
the Marie Curie Research Training Network. SG acknowledges the support of
NASA through the award of an Einstein Postdoctoral Fellowship. MG
acknowledges the financial contribution from contracts ASI-INAF I/023/05/0
and I/088/06/0 and Chandra grant GO0-11136X.

\bibliographystyle{apj}
\bibliography{../paper}


\end{document}